\documentclass[a4paper,twoside,twocolumn,english,aps,pra,showpacs,floats,eqsecnum,floatfix]{revtex4}
\usepackage[T1]{fontenc}
\usepackage[latin9]{inputenc}
\usepackage{amsmath}
\usepackage{graphicx}
\usepackage{amssymb}

\makeatletter

\providecommand{\tabularnewline}{\\}

\usepackage{babel}
\makeatother

\begin{document}

\title{Hybrid phase-space simulation method for interacting Bose fields}

\author{Scott E. Hoffmann, Joel F. Corney and Peter D. Drummond}

\address{ARC Centre for Quantum-Atom Optics, School of Physical Sciences,
University of Queensland, Brisbane, Queensland 4072, Australia}

\begin{abstract}
We introduce an approximate phase-space technique to simulate the
quantum dynamics of interacting bosons. With the future goal of treating
Bose-Einstein condensate systems, the method is designed for systems
with a natural separation into highly occupied (condensed) modes and
lightly occupied modes. The method self-consistently uses the Wigner
representation to treat highly occupied modes and the positive-P representation
for lightly occupied modes. In this method, truncation of higher-derivative
terms from the Fokker-Planck equation is usually necessary. However,
at least in the cases investigated here, the resulting systematic
error, over a finite time, vanishes in the limit of large Wigner occupation
numbers. We tested the method on a system of two interacting anharmonic
oscillators, with high and low occupations, respectively. The Hybrid
method successfully predicted atomic quadratures to a useful simulation
time 60 times longer than that of the positive-P method. The truncated
Wigner method also performed well in this test. For the prediction
of the correlation in a quantum nondemolition measurement scheme,
for this same system, the Hybrid method gave excellent agreement with
the exact result, while the truncated Wigner method showed a large
systematic error.
\end{abstract}


\maketitle

\section{Introduction}

The aim of this paper is to introduce a new, approximate, stochastic
phase-space method and to test it on some simple problems with interacting
Bose fields. A future goal of our research is to use the method to
simulate the dynamics of interacting Bose-Einstein condensates (BECs).
The method is, in fact, designed for BEC problems, since it relies
on the ability to make a meaningful separation of a multimode system
into highly occupied (condensed) modes and lightly occupied modes.
Hence our two-mode test cases will be constructed to have one highly
occupied mode $(N\gg1)$ and one lightly occupied mode $(N\lesssim1).$

Previous authors have developed formalisms in which condensed atoms
are treated in a different way to non-condensed atoms. Castin and
Dum \citep{Castin1998} studied the dynamics of Bose-Einstein condensates
at very low temperatures using a Bogoliubov \citep{Bogoliubov1947}
approach, in which the boson field operator is written as a sum of
condensate-mode terms and non-condensate-mode terms. Their treatment
deals with number eigenstates rather than coherent states. They obtain
results as an asymptotic expansion in the square root of the fraction
of non-condensed atoms. Gardiner and Zoller \citep{Gardiner1998d},
in the third of their series on quantum kinetic theory, consider a
stationary non-condensate band at a fixed temperature acting as a
reservoir to the dynamic condensate modes. They derive a master equation
for the condensate modes, using a number-conserving formalism. Dalton
\citep{Dalton2007} calculates quantum correlation functions for boson
field operators to use in the interpretation of double-well BEC interferometry
experiments. The approach is a phase-space method for a distribution
functional, in which the Wigner representation is used for the condensed
modes and the positive-P representation for the non-condensed modes.
Our method will be seen to be substantially different from these three
approaches.

Besides in BEC evolution and collision problems, other typical cases
where disparate occupation numbers exist would be in the quantum Brownian
motion of a small number of massive particles inside a BEC, or in
the collision of weak and strong coherent light pulses in a nonlinear
optical fibre. Hence we also consider these systems to be candidates
for the Hybrid method.

The foundations of this work are the stochastic phase-space methods
developed to simulate the quantum dynamics of systems with many degrees
of freedom. In particular we consider the Wigner-Moyal \citep{Wigner1932,Moyal1949}
approach, and the positive-P method \citep{Gilchrist1997,Gardiner2004}.
We will see that both methods have wide applicability, but are ultimately
limited in the parameter regimes on which they can be used. The Wigner-Moyal
method generally requires a truncation to be able to map to a stochastic
process. The resulting approximate theory typically fails to give
correct results when significant numbers of modes with small mode
occupation numbers are present \citep{Deuar2007}. 

The positive-P method is exact, but when applied to large multimode
problems can often be used only for limited simulation times before
very large sampling error renders it unusable. The longest useful
simulation times, for a given interaction strength, are for lightly
occupied modes \citep{Gilchrist1997}.

The new phase-space method to be introduced here is a combination
of the Wigner and positive-P methods. In this Hybrid method, as we
will call it, highly occupied modes are treated with the Wigner representation
while lightly occupied modes use the positive-P representation. A
truncation of higher-order derivative terms is usually needed, but
the resulting approximate method is expected to be valid (over finite
times) to within corrections of the order of the reciprocal of the
large occupation numbers.

The Wigner method is used in the regime where it is known to perform
best and produces most simplification of the stochastic differential
equations. The positive-P method is used on the modes that introduce
most error in the truncated Wigner method. This latter choice is also
designed to lengthen the useful simulation time.

In this paper we will summarize the properties of the two representations,
and discuss their successes and problems, before actual construction
of the Hybrid method. As a test case, we will apply the method to
an exactly solvable problem: a system of two coupled anharmonic oscillators,
one highly occupied, the other lightly occupied. The interaction preserves
individual particle numbers.

At first we simply calculate the expectation values of quadratures
and compare with the truncated Wigner method, the positive-P method
and the exact solution. Then we investigate a higher-order correlation
in the same system, one that would be observed in a quantum nondemolition
measurement (QND) scheme.

\section{The Single and Doubled Wigner Representations}

We consider a quantum many-body system of bosons. The relevant creation
and annihilation operators are denoted $\hat{a}_{m}^{\dagger}$, $\hat{a}_{m}$
. In the Wigner-Moyal approach, one complex phase space variable,
$\alpha_{m},$ is used for each mode, $m,$ of a system, and we call
this a single phase space. In contrast, a doubled phase space uses
two complex variables, $\alpha_{m}$ and $\alpha_{m}^{+},$ for each
mode. We will find that using the Wigner and positive-P representations
for different modes of the same system will generally require using
a doubled phase space, although this can be avoided in certain cases.

We begin by showing the definition and properties of the doubled Wigner
representation. This is an extension of the familiar single phase
space Wigner representation to a doubled phase space, and has been
studied and applied by Plimak \emph{et al} \citep{Plimak2001}. Throughout
this paper we will set $\hbar=1.$

The single phase-space Wigner representation of the density matrix
is given by\begin{eqnarray}
\hat{\rho} & = & \int d^{2}\alpha\, W(\alpha)\hat{\Lambda}_{W}(\alpha).\label{eq:W1}\end{eqnarray}
This is an expansion of the density matrix on a basis of operators,
the standard form we will use to compare all representations. Here
$W(\alpha)$ is the Wigner function on phase space. Following Moyal
\citep{Moyal1949} and Glauber and Cahill \citep{Cahill1969}, \begin{equation}
\hat{\Lambda}_{W}(\alpha)=\int\frac{d^{2}\xi}{\pi}\, e^{(\xi\hat{a}^{\dagger}-\xi^{*}\hat{a})}\, e^{(\alpha\xi^{*}-\alpha^{*}\xi)}\label{eq:W2}\end{equation}
is an operator function on phase space, with trace unity. We also
refer to this as the operator basis. We note that equation~(\ref{eq:W1})
has a unique inverse, defining the Wigner function in terms of the
density matrix:\begin{equation}
W(\alpha)=\int\frac{d^{2}\xi}{\pi^{2}}\, e^{(-\xi\alpha^{*}+\xi^{*}\alpha)}\,\mathrm{\mathnormal{\textrm{Tr}}}(\hat{\rho}e^{(\xi\hat{a}^{\dagger}-\xi^{*}\hat{a})}).\label{eq:W4}\end{equation}

By manipulating equation~(\ref{eq:W2}), these basis operators can
be written in the normally ordered Gaussian form of Corney and Drummond
\citep{Corney2003}, \begin{equation}
\hat{\Lambda}_{W}(\alpha)=2:e^{-2(\hat{a}^{\dagger}-\alpha^{*})(\hat{a}-\alpha)}:,\label{eq:W5}\end{equation}
where $:f(\hat{a},\hat{a}^{\dagger}):$ indicates normal ordering.

Now we may define the doubled Wigner representation with an expansion
of the density matrix of the form\begin{equation}
\hat{\rho}=\int d^{2}\alpha\int d^{2}\alpha^{+}\, W(\alpha,\alpha^{+})\hat{\Lambda}_{W}(\alpha,\alpha^{+}).\label{eq:W6}\end{equation}
Here $W(\alpha,\alpha^{+})$ is a Wigner function defined on a doubled
phase space and\begin{equation}
\hat{\Lambda}_{W}(\alpha,\alpha^{+})=2:e^{-2(\hat{a}^{\dagger}-\alpha^{+})(\hat{a}-\alpha)}:\label{eq:W7}\end{equation}
are the operator basis elements, also defined on the doubled phase
space. The new variable $\alpha^{+}$ appears where $\alpha^{*}$
had been, but in a stochastic simulation may take values different
from the complex conjugate of $\alpha.$

From equation~(\ref{eq:W7}) we can derive the \emph{operator correspondences}
for the doubled Wigner representation. The action of a creation or
annihilation operator, multiplying the density matrix to the left
or right, is equivalent to a linear differential operator acting on
the Wigner function:\begin{equation}
\hat{a}\hat{\rho}\leftrightarrow\left(\alpha+\frac{1}{2}\frac{\partial}{\partial\alpha^{+}}\right)\mathnormal{\mathcal{\mathrm{\mathnormal{W}}}}(\alpha,\alpha^{+})\label{eq:W8}\end{equation}
\begin{equation}
\hat{\rho}\hat{a}\leftrightarrow\left(\alpha-\frac{1}{2}\frac{\partial}{\partial\alpha^{+}}\right)\mathcal{\mathnormal{W}}(\alpha,\alpha^{+})\label{eq:W9}\end{equation}
\begin{equation}
\hat{a}^{\dagger}\hat{\rho}\leftrightarrow\left(\alpha^{+}-\frac{1}{2}\frac{\partial}{\partial\alpha}\right)\mathcal{\mathnormal{W}}(\alpha,\alpha^{+})\label{eq:W10}\end{equation}
\begin{equation}
\hat{\rho}\hat{a}^{\dagger}\leftrightarrow\left(\alpha^{+}+\frac{1}{2}\frac{\partial}{\partial\alpha}\right)\mathcal{\mathnormal{W}}(\alpha,\alpha^{+}).\label{eq:W11}\end{equation}
We add a cautionary note. The derivation of equations~(\ref{eq:W8}-\ref{eq:W11})
depends on the vanishing of boundary terms in an integration by parts.
This problem is discussed in Section IV.

We note that a pure coherent state (with $\rho=|\gamma\rangle\langle\gamma|$)
can be represented, in the doubled Wigner representation, with the
stochastic prescription\begin{equation}
\alpha=\gamma+\frac{1}{2}(n_{1}+in_{2})\label{eq:W12}\end{equation}
\begin{equation}
\alpha^{+}=\gamma^{*}+\frac{1}{2}(n_{1}-in_{2})\label{eq:W13}\end{equation}
where $n_{1}$ and $n_{2}$ are independent real Gaussian random noises
with unit standard deviations.

In this symmetrically ordered representation, the formula for estimating
symmetrically averaged products of creation and annihilation operators
as stochastic averages over trajectories is:\begin{equation}
\langle\hat{a}^{\dagger m}\hat{a}^{n}\rangle_{\mathrm{sym}}=\langle\langle\alpha^{+m}\alpha^{n}\rangle\rangle.\label{eq:W14}\end{equation}
We will use the notation $\langle\langle\rangle\rangle$ throughout
to signify a stochastic average over an ensemble of trajectories.

We note that we will be exploiting the \emph{nonuniqueness} of this
representation and that of the positive-P representation: an infinity
of different functions $W(\alpha,\alpha^{+})$ can give the same density
matrix according to equation~(\ref{eq:W6}). This feature of representations
on doubled phase spaces will allow us, in the case of the Hybrid representation,
to construct quasiprobabilities that are everywhere real and non-negative,
obeying Fokker-Planck equations that allow mapping to a stochastic
simulation.

\section{Problems with the Truncated Wigner method}

The (single phase space) Wigner representation has been widely used
to study diverse physical problems, with great success \citep{Graham1973a,Steel1998b,Sinatra2001}.
But a truncation of terms is necessary in most applications to allow
a stochastic simulation. The truncated Wigner method is then not exact.
Although the method gives good results in many cases, because the
truncation of terms can be well justified if mode occupation numbers
are large and simulation times are limited, the systematic errors
can be significant if those conditions are not met. In addition, the
estimation of a higher order moment (the expectation of a product
of more than one field operator) will generally contain a larger systematic
error than the estimation of the expectation value of a single field
operator \citep{Drummond1995}.

Even when no truncation is necessary, there is the problem of large
sampling error in a truncated Wigner simulation. While an initial
coherent state can be represented by a positive-P distribution of
zero width (see equation (\ref{eq:4.10})), the same state will have
a Wigner distribution with a finite width (equations (\ref{eq:W12}),
(\ref{eq:W13})). For short times, the growing positive-P noise will
not overtake the relatively constant Wigner noise. The result is greater
sampling error in the Wigner simulation, requiring the calculation
of far more trajectories to achieve the same precision.

The investigation of Deuar and Drummond \citep{Deuar2007} into BEC
scattering showed how the truncation problem produces serious systematic
errors in the simulation of a large number of interacting modes with
many lightly occupied modes. We will discuss these problems later
in this section.

Here we outline the reasons for truncation and the region of validity
of the approximation.

From equation~(\ref{eq:W4}) it may be seen that the Wigner function
is always real, but it may take negative values for some density matrices.
This would prevent us from mapping our quantum mechanics problem to
a stochastic simulation, since the latter would require a positive
semidefinite quasiprobability distribution.

However, when we find the equation of motion for the Wigner function,
the opportunity for an approximation procedure becomes apparent. This
equation follows from the operator correspondences of the single Wigner
representation (which can be obtained from equations~(\ref{eq:W8}-\ref{eq:W11})
with the replacement $\alpha^{+}\rightarrow\alpha^{*}$) and the evolution
equation for the density matrix\begin{equation}
\frac{\partial\hat{\rho}}{\partial t}=-i[\hat{H},\hat{\rho}].\label{eq:3.1}\end{equation}

We are going to restrict our attention to Hamiltonians, including
multimode Hamiltonians, that include products of creation and annihilation
operators only up to \emph{quartic} terms. This restriction will include
the model of BECs with two-body $s$-wave scattering \citep{Leggett2001}.
The equation for the evolution of a Wigner function under such a Hamiltonian
will always take the general form\begin{align}
\frac{\partial W}{\partial t} & =-\frac{\partial}{\partial\alpha}(A(\alpha)W(\alpha))-\frac{\partial}{\partial\alpha^{*}}(A^{*}(\alpha)W(\alpha))+T_{3}.\label{eq:3.2}\end{align}
Here $T_{3}$ is a term with three derivative operators, each either
$\frac{\partial}{\partial\alpha}$ or $\frac{\partial}{\partial\alpha^{*}}.$
The key point to note is that for undamped (unitary) time-evolution,
there are never any second order (diffusion) terms, which is a consequence
of the fact that the Wigner representation is symmetrically ordered.
Also, fourth-order terms always cancel. These general results for
quartic Hamiltonians, for the Wigner representation and for the positive-P
representation, are summarized in Table 1.

\noindent \begin{center}
\begin{table}[h]
\begin{centering}
\begin{tabular}{|c|c|c|c|}
\hline 
 & {\small Drift Terms} & {\small Diffusion Terms} & {\small Third-Order Terms}\tabularnewline
\hline
\hline 
{\small Wigner} & {\small Yes} & {\small No} & {\small Yes}\tabularnewline
\hline 
{\small Positive-P} & {\small Yes} & {\small Yes} & {\small No}\tabularnewline
\hline
\end{tabular}\smallskip{}

\par\end{centering}

\noindent \begin{raggedright}
Table 1: Terms in the Fokker-Planck equation for a quartic Hamiltonian,
using the Wigner and positive-P representations.
\par\end{raggedright}
\end{table}

\par\end{center}

It is found that the third order terms, including for more general
multimode problems, may be truncated and produce a systematic error
in $\frac{\partial W}{\partial t}$ that is relatively small compared
to the other terms, in the limit that the occupation numbers of the
modes remain very much greater than unity.

The motivation for this truncation is clear: equation~(\ref{eq:3.2})
then reduces to Liouville form, a special case of the Fokker-Planck
equation in which only drift terms influence the evolution of the
quasiprobability. If the initial density matrix for the problem is
such that the Wigner function is everywhere non-negative (and this
is a common situation) then the function will remain non-negative
for all times. A further mapping to a stochastic simulation becomes
possible. The only noise in the simulation will come from the initial
condition, since no second-order terms are present to cause diffusion.

A small error in $\frac{\partial W}{\partial t}$ will produce a large
error in $W$ after a sufficiently long time, so this approximation
procedure can only be valid for a finite time. Over the relevant time-scales,
the truncation is justified by a scaling argument. If the stochastic
variable $\alpha$ is seen from the truncated equations of motion
to remain of very large magnitude ($|\alpha|\sim\sqrt{N_{0}}\gg1$),
then we define a scaled variable $z=\alpha/\sqrt{N_{0}}$ and find
that the third-order terms take the form\begin{equation}
T_{3}\sim\frac{1}{N_{0}}\partial\partial\partial(\zeta W),\label{eq:3.3}\end{equation}
where $\partial$ is either $\frac{\partial}{\partial z}$ or $\frac{\partial}{\partial z^{*}}$
and $\zeta$ is either $z$ or $z^{*}.$

Deuar and Drummond \citep{Deuar2007} applied the truncated Wigner
method to the large multimode problem of scattering BECs and found
an ultraviolet divergence problem: systematic errors that grow with
the momentum cutoff imposed on the lattice. They were able to simulate
a BEC collision with 150,000 bosons, using the positive-P representation
for times long enough to obtain useful results, and thereby had an
exact result to compare with the truncated Wigner method. The latter
method produced a {}``false halo'' of particles in momentum space,
depletion leading to unphysical negative densities beyond the halo,
and accumulation of particles at low momenta - all in disagreement
with the exact positive-P results. The Wigner method requires that
initially empty modes of the system be represented by nonzero distributions,
as if one half of a virtual particle occupied each mode. Evidently
the truncated Wigner method treats these virtual particles as if they
were real, in that a scattering event involving them can produce real
populations of product modes.

This is an ultraviolet divergence problem in that it becomes worse
as the momentum cutoff is increased. To obtain the most physically
relevant results from a simulation, one must extrapolate to the continuum
limit. It is in this limit, as the momentum cutoff approaches infinity,
that the truncation errors are divergent. Clearly a full Wigner-Moyal
treatment without truncation would not have these errors, but such
a full theory with third-order derivatives also involves negative
probabilities, which have no stochastic equivalent.

We mention the projection method used with the truncated Wigner approach
\citep{Norrie2005,Norrie2006}, which amounts to another way to implement
a cutoff, but does not solve this ultraviolet divergence problem. 

We mention here the projection methods as other techniques (not exact)
for dealing with this problem .

This discussion of problems with the truncated Wigner method is given
as motivation for a Hybrid treatment. In future applications to multimode
systems, we will investigate whether the Hybrid method avoids these
problems. The large number of initial vacuum modes in a collision,
for example, would be treated in the Hybrid method with the positive-P
representation as phase-space variables set identically to zero. However
this requires a detailed future investigation. The problem is absent
in the pure positive-P method, although at long times very large sampling
errors are found instead \citep{Deuar2007}.

\section{The Positive-P method}

The positive-P method involves an extension of the Glauber-Sudarshan
P representation \citep{Glauber1963b,Sudarshan1963} from a single
phase space to a doubled phase space, the same procedure that gives
the doubled Wigner representation from the single. The defining equation
(for a single-mode problem) gives a representation of the density
matrix in terms of a $c$-number function, $P,$ and nondiagonal projection
operators, $\Lambda_{P},$ both defined on a doubled phase space:\begin{equation}
\hat{\rho}=\int d^{2}\alpha\int d^{2}\alpha^{+}\, P(\alpha,\alpha^{+})\hat{\Lambda}_{P}(\alpha,\alpha^{+}),\label{eq:4.1}\end{equation}
with\begin{equation}
\hat{\Lambda}_{P}(\alpha,\alpha^{+})=\frac{|\alpha\rangle\langle\alpha^{+*}|}{\langle\alpha^{+*}|\alpha\rangle}.\label{eq:4.2}\end{equation}

Here $\left|\alpha\right\rangle $ indicates a coherent state: a normalised
eigenstate of the annihilation operator $\hat{a}$. The effect of
left- and right- multiplication of the density matrix by $a$ and
$a^{\dagger}$ on $P(\alpha,\alpha^{+})$ can be deduced from Eqs.(\ref{eq:4.1},\ref{eq:4.2}).
The proof involves an integration by parts in which boundary terms
are assumed to vanish. The realm of validity of this assumption and
the resulting effects on stochastic simulations are discussed at length
by Gilchrist \emph{et al} \citep{Gilchrist1997}. When the boundary
terms vanish, the operator correspondences are:\begin{equation}
\hat{a}\hat{\rho}\leftrightarrow\alpha\mathnormal{\mathcal{\mathrm{\mathnormal{P}}}}(\alpha,\alpha^{+})\label{eq:4.3}\end{equation}
\begin{equation}
\hat{\rho}\hat{a}\leftrightarrow\left(\alpha-\frac{\partial}{\partial\alpha^{+}}\right)\mathnormal{\mathcal{\mathrm{\mathnormal{P}}}}(\alpha,\alpha^{+})\label{eq:4.4}\end{equation}
\begin{equation}
\hat{a}^{\dagger}\hat{\rho}\leftrightarrow\left(\alpha^{+}-\frac{\partial}{\partial\alpha}\right)\mathnormal{\mathcal{\mathrm{\mathnormal{P}}}}(\alpha,\alpha^{+})\label{eq:4.5}\end{equation}
\begin{equation}
\hat{\rho}\hat{a}^{\dagger}\leftrightarrow\alpha^{+}\mathnormal{\mathcal{\mathrm{\mathnormal{P}}}}(\alpha,\alpha^{+}).\label{eq:4.6}\end{equation}

As we noted in Table 1, all quartic Hamiltonian problems, in the positive-P
representation, give a true Fokker-Planck equation, with at most drift
and diffusion terms:\begin{equation}
\frac{\partial\mathnormal{\mathcal{\mathrm{\mathnormal{P}}}}}{\partial t}=-\partial_{\mu}(A^{\mu}(\alpha,\alpha^{+})\mathnormal{\mathcal{\mathrm{\mathnormal{P}}}})+\frac{1}{2}\partial_{\mu}\partial_{\nu}(D^{\mu\nu}(\alpha,\alpha^{+})P),\label{eq:4.7}\end{equation}
with\begin{equation}
\partial_{1}=\frac{\partial}{\partial\alpha},\ \partial_{2}=\frac{\partial}{\partial\alpha^{+}},\label{eq:4.8}\end{equation}
and summation over $\mu,\nu$ implied. Thus, for the many-boson Hamiltonian
with two-body $s-$wave scattering terms \citep{Leggett2001}, no
truncation of the positive-P equations is needed.

The positive-P method solves two problems that occur with single phase-space
representations. First, if the distribution function is not guaranteed
to remain real and non-negative, we cannot map the dynamics onto a
stochastic simulation using standard methods. To deal with this, we
use the feature of the representation that an infinity of different
functions, $P(\alpha,\alpha^{+}),$ may represent the same density
matrix. We may choose, for the initial condition, a particular function

\begin{equation}
P_{+}(\alpha,\alpha^{+})=\frac{1}{4\pi^{2}}e^{-\frac{1}{4}|\alpha-\alpha^{+*}|^{2}}\langle\frac{1}{2}(\alpha+\alpha^{+*})|\hat{\rho}|\frac{1}{2}(\alpha+\alpha^{+*})\rangle,\label{eq:4.9}\end{equation}
that satisfies Eq.(\ref{eq:4.1}) and is everywhere non-negative,
as required. Alternatively, an initial pure coherent state, with $\hat{\rho}=|\gamma\rangle\langle\gamma|$,
can have a delta function representation, also positive:\begin{equation}
\mathnormal{\mathcal{\mathrm{\mathnormal{P}}}}(\alpha,\alpha^{+})=\delta^{2}(\alpha-\gamma)\delta^{2}(\alpha^{+}-\gamma^{*}).\label{eq:4.10}\end{equation}
The stochastic representation of this initial condition is simply\begin{equation}
\alpha=\gamma,\quad\alpha^{+}=\gamma^{*}.\label{eq:4.11}\end{equation}

The second problem to deal with is that the diffusion matrix may not
be positive semidefinite when written in the basis of real ($x$)
and imaginary ($y$) parts $(\alpha_{x},\alpha_{x}^{+},\alpha_{y},\alpha_{y}^{+}).$
However, there is another symmetry in the positive-P representation,
arising from the analyticity in $\alpha$ and $\alpha^{+}$ of the
nondiagonal projection operator (\ref{eq:4.2}), that lets us make
replacements to the real and imaginary parts of the derivative operators
(\ref{eq:4.8}) in (\ref{eq:4.3}-\ref{eq:4.6}), in just such a way
that the resulting Fokker-Planck equation, relative to the component
basis, has a positive semidefinite diffusion matrix \citep{Gardiner2004}.
With a positive initial condition and a true, positive semidefinite
Fokker-Planck equation, the distribution is guaranteed to remain positive.
The standard method of mapping to a stochastic simulation also requires
a diffusion matrix that is positive semidefinite (all of its eigenvalues
are non-negative), so that stochastic equations are immediately derivable. 

The final step of mapping to stochastic differential equations involves
first finding an $N$-noise factorization of the diffusion matrix
of the form\begin{equation}
D^{\mu\nu}=\sum_{n=1}^{N}B^{\mu n}B^{\nu n}.\label{eq:4.12}\end{equation}
This introduces another gauge degree of freedom that we will exploit
later. Different choices of the factor matrix, $B,$ that satisfy
(\ref{eq:4.12}) may provide stochastic simulations with widely different
sampling error characteristics.

The result of the adjustment of the diffusion matrix and this choice
of the factor matrix is the set of It\^o stochastic differential
equations\begin{equation}
d\alpha^{\mu}=A^{\mu}dt+\sum_{n=1}^{N}B^{\mu n}dw_{n},\label{eq:4.13}\end{equation}
where the $dw_{n}$ are $N$ real Weiner increments \citep{Gardiner2004}
satisfying the stochastic average\begin{equation}
\langle\langle dw_{n}(t)dw_{m}(t)\rangle\rangle=\delta_{nm}dt.\label{eq:4.14}\end{equation}
These SDEs, with appropriate initial conditions (equations (\ref{eq:4.11})
for coherent states), are used to evolve a large ensemble of trajectories.
The positive-P representation is normally ordered, meaning that the
most easily calculated quantum mechanical expectation values are of
normally ordered operators. The formula for estimating a normally
ordered quantum mechanical expectation value as a stochastic average
is:\begin{equation}
\langle\hat{a}^{\dagger m}\hat{a}^{n}\rangle=\langle\langle\alpha^{+m}\alpha^{n}\rangle\rangle.\label{eq:4.in}\end{equation}
We see from equation (\ref{eq:4.11}) that a coherent state can be
represented initially with no noise in the positive-P representation.
In this paper we will not embark on a detailed comparison of sampling
error in the truncated Wigner, positive-P and Hybrid methods. However,
we will take note of the number of trajectories needed, in each method,
for an ensemble average to converge to a satisfactory result. All
of our simulations were performed using xmds \citep{xmds}, and we
used the built-in sampling error estimates of that program to judge
convergence.

The next part of our construction of the Hybrid representation involves
writing the nondiagonal projectors for the positive-P representation
in normally-ordered Gaussian form. The result of manipulating equation
(\ref{eq:4.2}) is \begin{equation}
\Lambda_{P}(\alpha,\alpha^{+})=:e^{-(\hat{a}^{\dagger}-\alpha^{+})(\hat{a}-\alpha)}:.\label{eq:4.15}\end{equation}

\section{Problems with the Positive-P Method}

A particular choice of the factor matrix, $B,$ gives a set of stochastic
differential equations (\ref{eq:4.13}) that governs the evolution
of the ensemble of trajectories. Unless this evolution is constrained
in some way, trajectories may wander far from each other in phase
space. Then the averaging over trajectories to estimate an expectation
value may involve additions of many different, extremely large, numbers.
No computer can calculate such an average without incurring a very
large roundoff error.

The result is the dramatic rise in sampling error that has been seen
in some positive-P simulations. The growth in width of the distribution
of trajectories often occurs over a short time scale, so that the
sampling error suddenly rises by many orders of magnitude, with the
resulting growth of numerical errors. The simulation is of no value
beyond this critical time.

The problem can be caused by drift terms or noise terms, or a combination
of both. A single-mode example to illustrate these problems is the
anharmonic oscillator, with Hamiltonian\begin{equation}
\hat{H}=\omega\hat{a}^{\dagger}\hat{a}+\chi\hat{a}^{\dagger}\hat{a}^{\dagger}\hat{a}\hat{a}\label{eq:5.0}\end{equation}
and positive-P It\^o stochastic differential equations:\begin{equation}
d\alpha=-i(\omega+2\chi\alpha^{+}\alpha)\alpha dt+\sqrt{-2i\chi}dw\label{eq:5.1}\end{equation}
\begin{equation}
d\alpha^{+}=+i(\omega+2\chi\alpha^{+}\alpha)\alpha^{+}dt+\sqrt{2i\chi}dw^{+}.\label{eq:5.2}\end{equation}
If we ignore the noise terms and choose $\alpha^{+}\alpha$ as a real
number initially, the trajectory will be a circle in the complex $\alpha$
plane. But if $\alpha^{+}\alpha$ includes an imaginary part (from
noise or from an initial condition other than the coherent state condition
(\ref{eq:4.11})), either $\alpha$ or $\alpha^{+}$ will spiral towards
infinity, while the other spirals in towards the origin. The noise
terms of the SDEs contribute to the problem, since they generally
move $\alpha^{+}\alpha$ away from real values, thereby inducing the
spiraling.

Note that the single Wigner representation does not suffer from this
problem because the real term $|\alpha|^{2}$ will always appear in
place of $\alpha^{+}\alpha$ in the SDEs.

Sampling error growth can be reduced or postponed by using our freedom
to choose different factor matrices that give the same diffusion matrix,
and by modifying the drift equations. Such methods are called stochastic
gauge techniques \citep{Deuar2002,Deuar2003,Dowling2005}. However,
while these are useful in single-mode examples, they are somewhat
complicated when generalized to multi-mode cases. Also, we are interested
in extending the time available for useful, error-free simulations
to even longer time-scales than these methods can provide.

A special case of problems with drift trajectories is when a trajectory
is capable of reaching infinity in a finite time \citep{Gilchrist1997}.
This typically results in power-law tails in the distribution function,
which violates the assumption that partial integration can be carried
out. The simulation as it stands is then invalid beyond the singularity
time. This problem can be dealt with using drift gauges \citep{Deuar2002}.
Our examples will not fit into this category.

We simulated the anharmonic oscillator in the positive-P representation
to illustrate the sampling error problem. Figure 1 shows the $X$
quadrature ($\hat{X}=\frac{1}{2}(\hat{a}+\hat{a}^{\dagger}))$, with
the choices $\omega=0$ (for simplicity) and initial average number
$N_{0}=1.$ As we will do for all of our simulations, we plot results
against a scaled time parameter, in this case $\chi t$. (This is
dimensionless when using $\hbar=1$).

\begin{center}
\begin{figure}[h]
\begin{centering}
\includegraphics[scale=0.5]{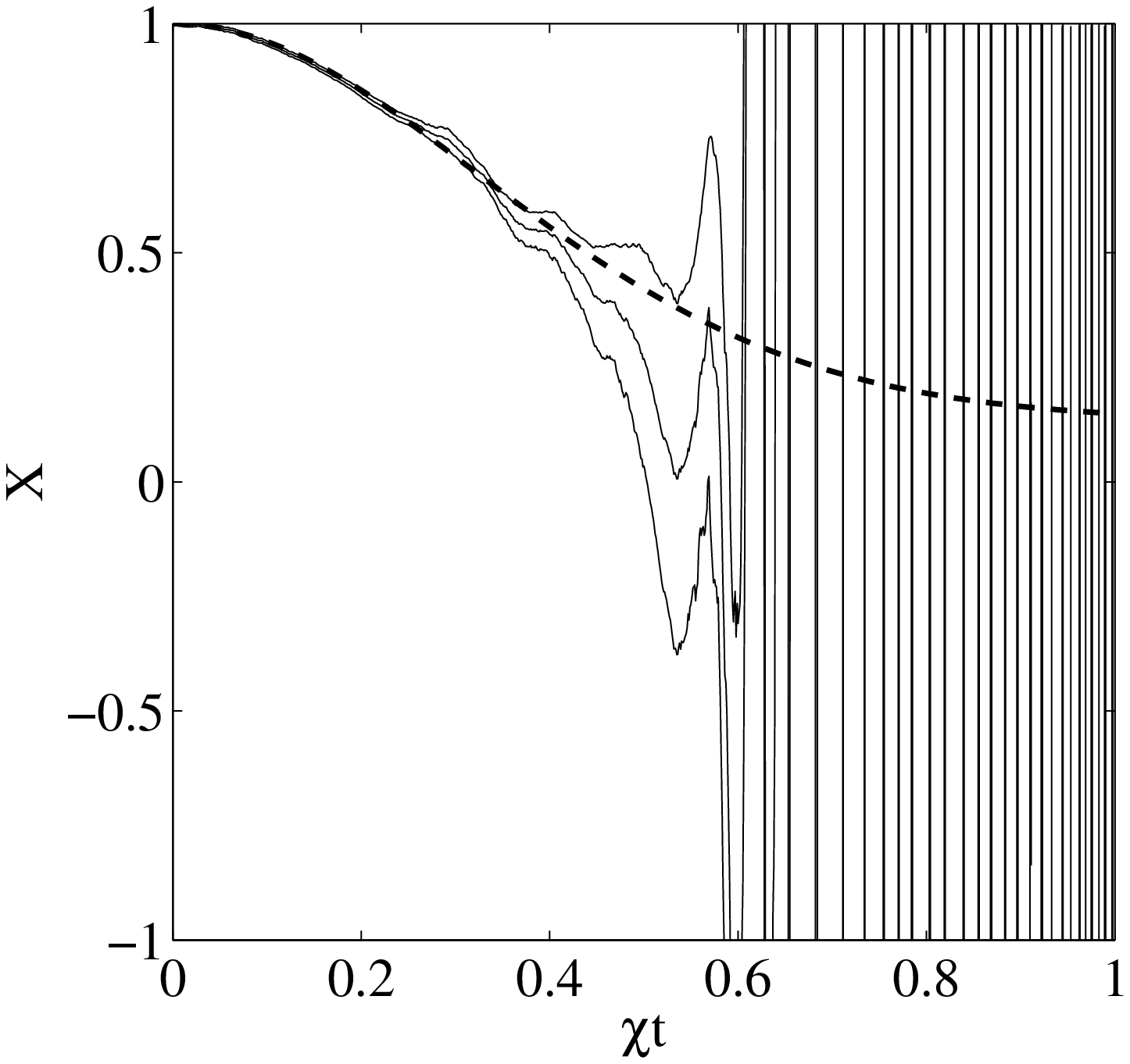}
\par\end{centering}

\noindent \begin{raggedright}
FIG. 1. X quadrature for single mode anharmonic oscillator \textit{vs}
$\chi t$: positive-P method. Plotted are the ensemble average and
the ensemble average~$\pm$~sampling error estimate. The dashed
line is the exact result. Parameters: $\omega=0$, $N_{0}=1.$ Number
of trajectories: 1,000. 
\par\end{raggedright}

\end{figure}

\par\end{center}

Deuar and Drummond \citep{Deuar2002} have investigated various factors
that affect the time for sampling error to become unmanageable in
a multimode positive-P simulation. They have found that coarser spatial
lattices, weaker interactions and lower particle densities all extend
the lifetime of the simulation. Of course the spatial lattice spacing
can only be increased at the expense of systematic error, while the
other two factors are fixed by the system being simulated.

In the Hybrid scheme we will be using the positive-P representation
only for the modes with lowest occupations. Our test cases will investigate
whether this delays the onset of large sampling error.

\section{The Hybrid method}

The Hybrid method is designed to exploit a particular feature of Bose-Einstein
condensate systems: that a limited number of modes have very high
occupation numbers. The method involves separating the physical system
into modes that are, at least initially, highly occupied (the condensed
modes) and those that are lightly occupied (the output of an atom
laser or the products of a BEC collision). Then we intend to use different
representations to treat different modes, treating the highly occupied
modes with a form of the Wigner representation and the lightly occupied
modes with the positive-P representation.

Use of the Wigner representation for the highly occupied modes will
in general simplify the structure of the resulting diffusion matrix.
In a simple two-mode model discussed in Section VII, we will see that
this allows us to delay the rapid growth of sampling error. And by
\emph{not} using the Wigner representation for the potentially very
large number of lightly occupied modes, we intend to avoid the false
halo problem.

Our first task is to show that we can consistently use two different
representations on different modes, correctly describing interactions
that couple these different modes. For general interactions of this
sort, diffusion terms involving the Wigner modes are inevitable. To
be able to construct a positive semidefinite diffusion matrix in the
general case, we will have to use a doubled phase space throughout.

Now we can exploit the similarities in equations (\ref{eq:W7}) and
(\ref{eq:4.15}) to define a Hybrid representation with a particularly
simple notation. We suppose that a system has modes labelled $m=1,\dots,M$.
These modes are to be treated with the Wigner representation or the
positive-P representation depending on whether a parameter $r_{m}$
takes the value:\begin{equation}
r_{m}=1\quad\mbox{for positive-P,}\label{eq:6.1}\end{equation}
 \begin{equation}
r_{m}=2\quad\mbox{for Wigner.}\label{eq:6.2}\end{equation}
The nondiagonal projection operator is a direct product of terms for
each mode, $m$:\begin{equation}
\hat{\Lambda}_{H}(\vec{\alpha};\vec{r})=\prod_{m=1}^{M}r_{m}:e^{-r_{m}(\hat{a}^{\dagger}-\alpha^{+})(\hat{a}-\alpha)}:.\label{eq:6.3}\end{equation}
Then the Hybrid representation of the density matrix becomes\begin{equation}
\hat{\rho}=\int d^{4M}\vec{\alpha}\, P_{H}(\vec{\alpha};\vec{r})\hat{\Lambda}_{H}(\vec{\alpha};\vec{r}),\label{eq:6.4}\end{equation}
where $\vec{\alpha}=(\alpha_{1},\alpha_{1}^{+},\dots,\alpha_{M},\alpha_{M}^{+})$
and $\vec{r}=(r_{1},\dots,r_{M}).$

Note that the use of the parameter $r$ for these doubled phase space
representations is very much like Glauber and Cahill's use of the
parameter $s$ to span antinormally ordered, symmetrically ordered
and normally ordered single phase space representations. The connection
between the two schemes follows by taking\begin{equation}
r=\frac{s-1}{s}\label{eq:6.5}\end{equation}
 and mapping doubled to single phase spaces.

The applicability of the positive-P method depends on the two results
that we mentioned in Section IV. First, for any initial density matrix,
it is possible to choose a phase-space distribution that is everywhere
real and non-negative (using equation (\ref{eq:4.9})). Second, it
is always possible to cast any diffusion matrix into a form that is
equivalent with respect to physical predictions and that is positive
semi-definite in the basis of real and imaginary parts of the phase-space
variables. Corresponding results must hold for any Hybrid representation
in order for that method to be usable. We were able to prove both
assertions.

First, we found an integral transform that takes a positive-P distribution
to a doubled Wigner distribution representing the same density matrix.
We show the single-mode case: \begin{equation}
W(\alpha,\alpha^{+})=\frac{1}{2\pi}\int d^{2}\psi'\, e^{-\frac{1}{2}|\psi-\psi'|^{2}}\, P(\frac{1}{2}(\psi'+\chi),\frac{1}{2}(\psi'-\chi)),\label{eq:6.a}\end{equation}
with\begin{equation}
\psi=\alpha+\alpha^{+*},\quad\chi=\alpha-\alpha^{+*}.\label{eq:6.b}\end{equation}
Note that $P$ has four independent real parameters, but the integration
is over only two degrees of freedom. The single-mode case is shown
but the extension to the multi-mode case is straightforward. Since
the kernel is positive, the transform can be used to take the initial
positive distribution (\ref{eq:4.9}) to an everywhere positive doubled
Wigner distribution. Extension to the case with many modes treated
by different representations proves the first assertion.

The proof of the second assertion is exactly like the textbook proof
for the positive-P representation, since the derivative equivalences\begin{equation}
\frac{\partial}{\partial\alpha}\leftrightarrow\frac{\partial}{\partial\alpha_{x}}\leftrightarrow-i\frac{\partial}{\partial\alpha_{y}},\label{eq:6c}\end{equation}
\begin{equation}
\frac{\partial}{\partial\alpha^{+}}\leftrightarrow\frac{\partial}{\partial\alpha_{x}^{+}}\leftrightarrow-i\frac{\partial}{\partial\alpha_{y}^{+}},\label{eq:6d}\end{equation}
are the same as their positive-P counterparts.

Use of the Hybrid method is simple for few-mode problems. For the
mapping of the evolution equation (\ref{eq:3.1}) for $\rho$ to a
Fokker-Planck equation, we use either the Wigner (\ref{eq:W8}-\ref{eq:W11})
or positive-P (\ref{eq:4.3}-\ref{eq:4.6}) operator correspondences
as appropriate for each mode. In general there will be terms with
three derivative operators for quartic Hamiltonians. (Terms with four
derivatives always cancel.) For each application, we must decide whether
truncation of these terms, to produce a drift/diffusion problem, is
valid. Scaling arguments like those applied to the Wigner method (\ref{eq:3.3})
can be used here. In problems involving both highly occupied modes
and lightly occupied modes, there may occur problematic three-derivative
terms from mutual interaction of those modes. 

A feature peculiar to the Hybrid method is that there will appear
what we call \emph{interface noise}: there will be diffusion terms
that are proportional to the difference of $r$ values for different
modes, that would vanish if those modes were treated with the same
representation.

The mapping to stochastic differential equations uses the same rule
as is used for the positive-P representation: if a generally complex
matrix $B$ provides a factorization $D=BB^{T}$ of the diffusion
matrix, then the It\^o stochastic differential equations can be chosen
as\begin{equation}
d\alpha^{\mu}=A^{\mu}dt+\sum_{n=1}^{N}B^{\mu n}dw_{n},\label{eq:6.6}\end{equation}
where $\mu$ labels the components of the vector of phase space variables
$\vec{\alpha}=(\alpha_{1},\alpha_{1}^{+},\dots,\alpha_{M},\alpha_{M}^{+})$
and the $dw_{n}$ are $N$ real, independent Weiner increments. We
note that the freedom of choice of a factor matrix, $B,$ introduces
a gauge degree of freedom that may allow us to reduce sampling error
in simulations.

The relation between physical expectation values and stochastic averages
will take new forms in the Hybrid representation. Here an observable
may be a product of factors to be treated with the symmetrically ordered
Wigner representation and others to be treated with the normally ordered
positive-P representation. So, for example, in Section IX we will
need to calculate an expectation value as\begin{equation}
\langle\hat{N}_{a}\hat{Y}_{b}\rangle=\langle\frac{1}{2i}\hat{a}^{\dagger}\hat{a}(\hat{b}-\hat{b^{\dagger}})\rangle=\langle\langle\frac{1}{2i}(\alpha^{+}\alpha-\frac{1}{2})(\beta-\beta^{\dagger})\rangle\rangle,\label{eq:6.7}\end{equation}
where the $a$ mode is treated with the Wigner representation while
the $b$ mode is treated with the positive-P representation.

\section{Test Case: Coupled Anharmonic Oscillators}

As a first test of the Hybrid method, we simulated the behavior of
two coupled anharmonic oscillators, with a coupling that preserves
the individual mode occupations. The Hamiltonian is\begin{equation}
\hat{H}=\omega_{a}\hat{a}^{\dagger}\hat{a}+\chi_{a}\hat{a}^{\dagger}\hat{a}^{\dagger}\hat{a}\hat{a}+\omega_{b}\hat{b}^{\dagger}\hat{b}+\chi_{b}\hat{b}^{\dagger}\hat{b}^{\dagger}\hat{b}\hat{b}+g\hat{a}^{\dagger}\hat{a}\hat{b}^{\dagger}\hat{b}.\label{eq:7.1}\end{equation}
We used an initial coherent state for the $a$ mode (with high mean
occupation $N_{a0}=100$) and for the $b$ mode (low mean occupation
$N_{b0}=0.01$). We set $\omega_{a}=\omega_{b}=0$ for convenience
and used $\chi_{a}=\chi_{b}=g=1,$ which sets the scale for the time
variable.

This model is meant to resemble just a few terms of the much larger
multimode Hamiltonian for a Bose gas with $s$-wave scattering terms.

Note that we have chosen a model system in which the $a$ occupation
remains constantly large, while the $b$ occupation stays small. The
Hybrid method can be used in cases where these numbers are not conserved,
and gives good results when the occupations of the modes remain high
and low over the interaction time, respectively. Results from this
category will be presented in a later work.

We simulated this system using the Hybrid method and, for comparison,
the truncated Wigner method and the positive-P method. We were also
able to obtain an exact solution for coherent state initial conditions,
as did Chaturvedi and Srinivasan \citep{Chaturvedi1991}.

We insert the Hybrid representation (\ref{eq:6.4}) of the density
matrix into the evolution equation (\ref{eq:3.1}). An integration
by parts, justified in this case, amounts to using the Hybrid operator
correspondences (\ref{eq:W8}-\ref{eq:W11}, \ref{eq:4.3}-\ref{eq:4.6}).
This gives an equation of the form\begin{equation}
\int d^{8}\vec{\alpha}\frac{\partial P_{H}}{\partial t}\hat{\Lambda}_{H}(\vec{\alpha},\vec{r})=\int d^{8}\vec{\alpha}\mathcal{L}(\vec{\alpha},\vec{r})P_{H}(\vec{\alpha},\vec{r})\hat{\Lambda}_{H}(\vec{\alpha},\vec{r}),\label{eq:7.1a}\end{equation}
where $\mathcal{L}$ is a linear, differential operator that acts
on $P_{H}.$

We extract a Fokker-Planck equation for $P_{H}$ from (\ref{eq:7.1a}),
keeping all terms, including third-order derivative terms. (We note
that for doubled phase-space representations this choice is not unique.).
We find \begin{eqnarray}
i\frac{\partial P_{H}}{\partial t} & = & -\frac{\partial}{\partial\alpha}\{2\chi_{a}(\alpha^{+}\alpha-1)+g\beta^{+}\beta\}\alpha P_{H}\nonumber \\
 &  & +\frac{\partial}{\partial\alpha^{+}}\{2\chi_{a}(\alpha^{+}\alpha-1)+g\beta^{+}\beta\}\alpha^{+}P_{H}\nonumber \\
 &  & -\frac{\partial}{\partial\beta}\{2\chi_{b}\beta^{+}\beta+g(\alpha^{+}\alpha-\frac{1}{2})\}\beta P_{H}\nonumber \\
 &  & +\frac{\partial}{\partial\beta^{+}}\{2\chi_{b}\beta^{+}\beta+g(\alpha^{+}\alpha-\frac{1}{2})\}\beta^{+}P_{H}\nonumber \\
 &  & +\chi_{b}\frac{\partial^{2}}{\partial\beta^{2}}\beta^{2}P_{H}-\chi_{b}\frac{\partial^{2}}{\partial\beta^{+2}}\beta^{+2}P_{H}\nonumber \\
 &  & +\frac{g}{2}\frac{\partial}{\partial\alpha}\frac{\partial}{\partial\beta}\alpha\beta P_{H}+\frac{g}{2}\frac{\partial}{\partial\alpha}\frac{\partial}{\partial\beta^{+}}\alpha\beta^{+}P_{H}\nonumber \\
 &  & -\frac{g}{2}\frac{\partial}{\partial\alpha^{+}}\frac{\partial}{\partial\beta}\alpha^{+}\beta P_{H}-\frac{g}{2}\frac{\partial}{\partial\alpha^{+}}\frac{\partial}{\partial\beta^{+}}\alpha^{+}\beta^{+}P_{H}\nonumber \\
 &  & +\frac{g}{4}\frac{\partial}{\partial\alpha}\frac{\partial}{\partial\alpha^{+}}\{\frac{\partial}{\partial\beta}\beta-\frac{\partial}{\partial\beta^{+}}\beta^{+}\}P_{H}\nonumber \\
 &  & +\frac{\chi_{a}}{2}\{\frac{\partial^{2}}{\partial\alpha^{2}}\frac{\partial}{\partial\alpha^{+}}\alpha-\frac{\partial^{2}}{\partial\alpha^{+2}}\frac{\partial}{\partial\alpha}\alpha^{+}\}P_{H}.\label{eq:7.1.1}\end{eqnarray}
We use the convention that the derivative operators act on all factors
to the right.

To apply the scaling argument discussed in Section III, we write the
third-order derivative terms above (which we call $T_{3}$) in terms
of the scaled phase-space variables \begin{equation}
u\equiv\alpha/\sqrt{N_{a0}},\quad\quad u^{+}\equiv\alpha^{+}/\sqrt{N_{a0}},\label{eq:7.1b}\end{equation}
\begin{equation}
v\equiv\beta/\sqrt{N_{b0}},\quad\quad v^{+}\equiv\beta^{+}/\sqrt{N_{b0}}.\label{eq:7.1b.1}\end{equation}
Then $T_{3}$ becomes\begin{eqnarray}
T_{3} & = & \frac{1}{N_{a0}}\frac{g}{4}\frac{\partial}{\partial u}\frac{\partial}{\partial u^{+}}\{\frac{\partial}{\partial v}v-\frac{\partial}{\partial v^{+}}v^{+}\}P_{H}\nonumber \\
 &  & +\frac{1}{N_{a0}}\frac{\chi_{a}}{2}\{\frac{\partial^{2}}{\partial u^{2}}\frac{\partial}{\partial u^{+}}u-\frac{\partial^{2}}{\partial u^{+2}}\frac{\partial}{\partial u}u^{+}\}P_{H}.\label{eq:7.1c}\end{eqnarray}

We expect, in the stochastic simulation of this problem, that there
will be a finite time scale over which $\alpha$ and $\alpha^{+}$
will remain distributed close to order $\sqrt{N_{a0}}$ in magnitude,
while $\beta$ and $\beta^{+}$ remain near $\sqrt{N_{b0}}$. Our
first simulation will stay within this time region. Over that time
scale, the third-order derivative terms will make a negligible contribution
to $P_{H}$ compared to the drift terms (first-order terms which scale
like $N_{a0}$) and the diffusion terms (second-order terms which
scale like 1).

After we truncate these terms, the Fokker-Planck equation has drift
vector (in the basis $(\alpha,\alpha^{+},\beta,\beta^{+})$)\begin{equation}
A=\left(\begin{array}{c}
-i\{2\chi_{a}(\alpha^{+}\alpha-1)+g\beta^{+}\beta\}\alpha\\
+i\{2\chi_{a}(\alpha^{+}\alpha-1)+g\beta^{+}\beta\}\alpha^{+}\\
-i\{2\chi_{b}\beta^{+}\beta+g(\alpha^{+}\alpha-\frac{1}{2})\}\beta\\
+i\{2\chi_{b}\beta^{+}\beta+g(\alpha^{+}\alpha-\frac{1}{2})\}\beta^{+}\end{array}\right)\label{eq:7.2}\end{equation}
and diffusion matrix: \begin{align}
D & =\left(\begin{array}{cccc}
0 & 0 & -\frac{ig}{2}\alpha\beta & -\frac{ig}{2}\alpha\beta^{+}\\
0 & 0 & +\frac{ig}{2}\alpha^{+}\beta & +\frac{ig}{2}\alpha^{+}\beta^{+}\\
-\frac{ig}{2}\alpha\beta & +\frac{ig}{2}\alpha^{+}\beta & -2i\chi_{b}\beta^{2} & 0\\
-\frac{ig}{2}\alpha\beta^{+} & +\frac{ig}{2}\alpha^{+}\beta^{+} & 0 & +2i\chi_{b}\beta^{+2}\end{array}\right).\label{eq:7.3}\end{align}

Because of the use of the Wigner representation for the $a$ mode,
this diffusion matrix differs from the one resulting from a pure positive-P
treatment in the absence of terms $-2i\chi_{a}\alpha^{2}$ and $+2i\chi_{a}\alpha^{+2}$
in the first two diagonal spaces, respectively.

We were able to construct a factorization of the diffusion matrix
(\ref{eq:7.3}) by first treating the diagonal terms and then recognizing
a simple structure in the remaining matrix. The following factor matrix
requires only four real noises in the SDEs:\begin{eqnarray}
B & = & \sqrt{2i\chi_{b}}\left(\begin{array}{cccc}
0 & 0 & 0 & 0\\
0 & 0 & 0 & 0\\
i\beta & 0 & 0 & 0\\
0 & \beta^{+} & 0 & 0\end{array}\right)\nonumber \\
 &  & \quad+\frac{1}{2}\sqrt{-ig}\left(\begin{array}{cccc}
0 & 0 & \alpha & i\alpha\\
0 & 0 & -\alpha^{+} & -i\alpha^{+}\\
0 & 0 & \beta & -i\beta\\
0 & 0 & \beta^{+} & -i\beta^{+}\end{array}\right).\label{eq:7.4}\end{eqnarray}

The resulting SDEs produced the results shown in Figures 2 and 3.
We calculated the expectation values of the quadrature operators $\hat{X}_{a}=\frac{1}{2}(\hat{a}+\hat{a}^{\dagger})$
and $\hat{X}_{b}=\frac{1}{2}(\hat{b}+\hat{b}^{\dagger}).$ The simulation
was clearly stable over the time scale shown and gave results in excellent
agreement with the exact solution. We will refer to the method used
here as a gauge Hybrid method, since it relies on being able to find
a diffusion gauge (a useful factorization of the diffusion matrix).

\noindent \begin{center}
\begin{figure}[h]
\begin{centering}
\includegraphics[scale=0.4]{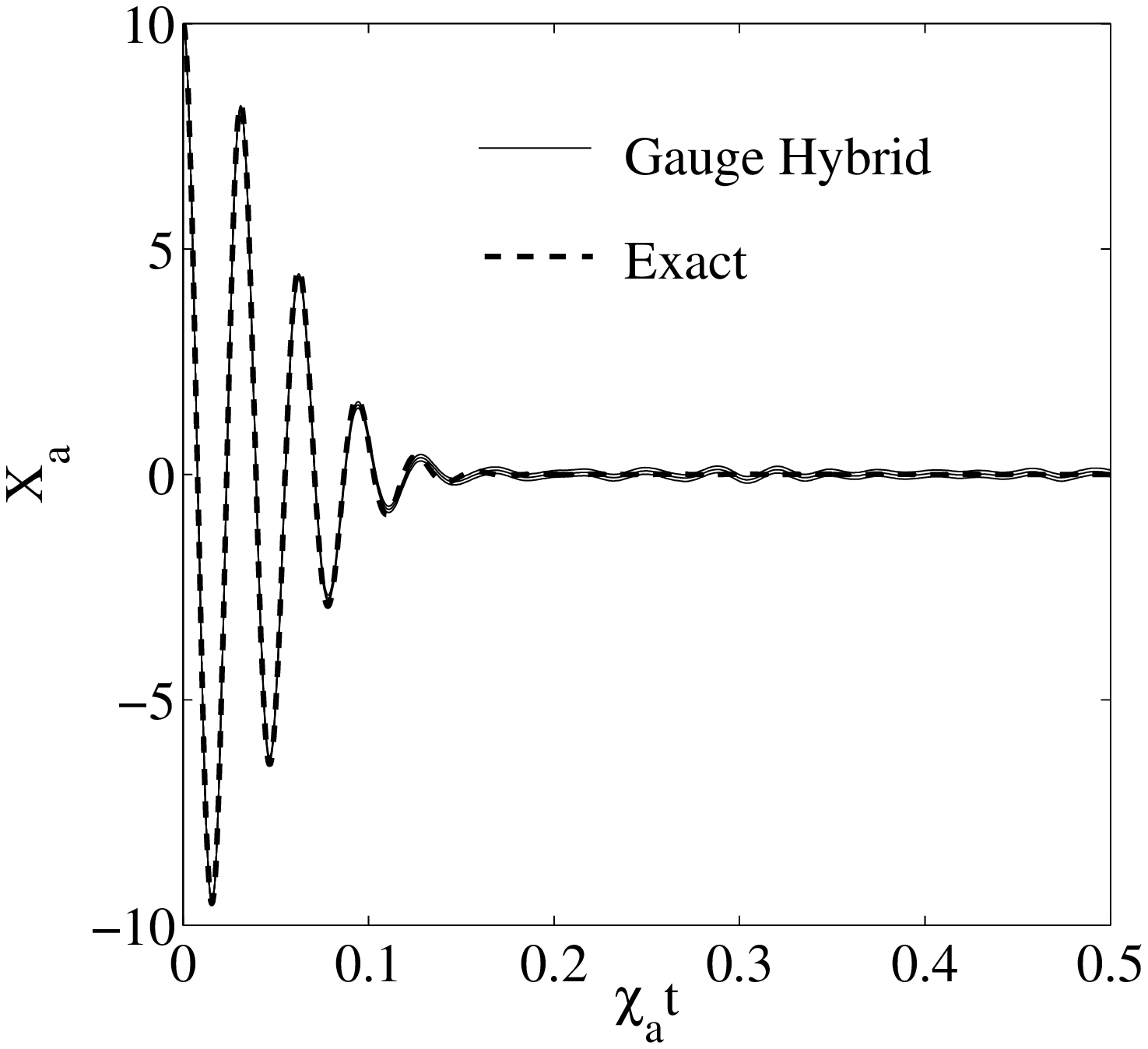}
\par\end{centering}

\noindent \begin{raggedright}
FIG. 2. X quadrature for mode $a$ \textit{vs} $\chi_{a}t$: coupled
anharmonic oscillators treated with the gauge Hybrid method. Plotted
are the ensemble average, the ensemble average~$\pm$~sampling error
estimate and the exact solution. Parameters: $\omega_{a}=\omega_{b}=0$,
$\chi_{a}=\chi_{b}=g,$ $N_{a0}=100$, $N_{b0}=0.01$. Number of trajectories:
10,000.
\par\end{raggedright}

\end{figure}

\par\end{center}

\noindent \begin{center}
\begin{figure}[h]
\noindent \begin{centering}
\includegraphics[scale=0.4]{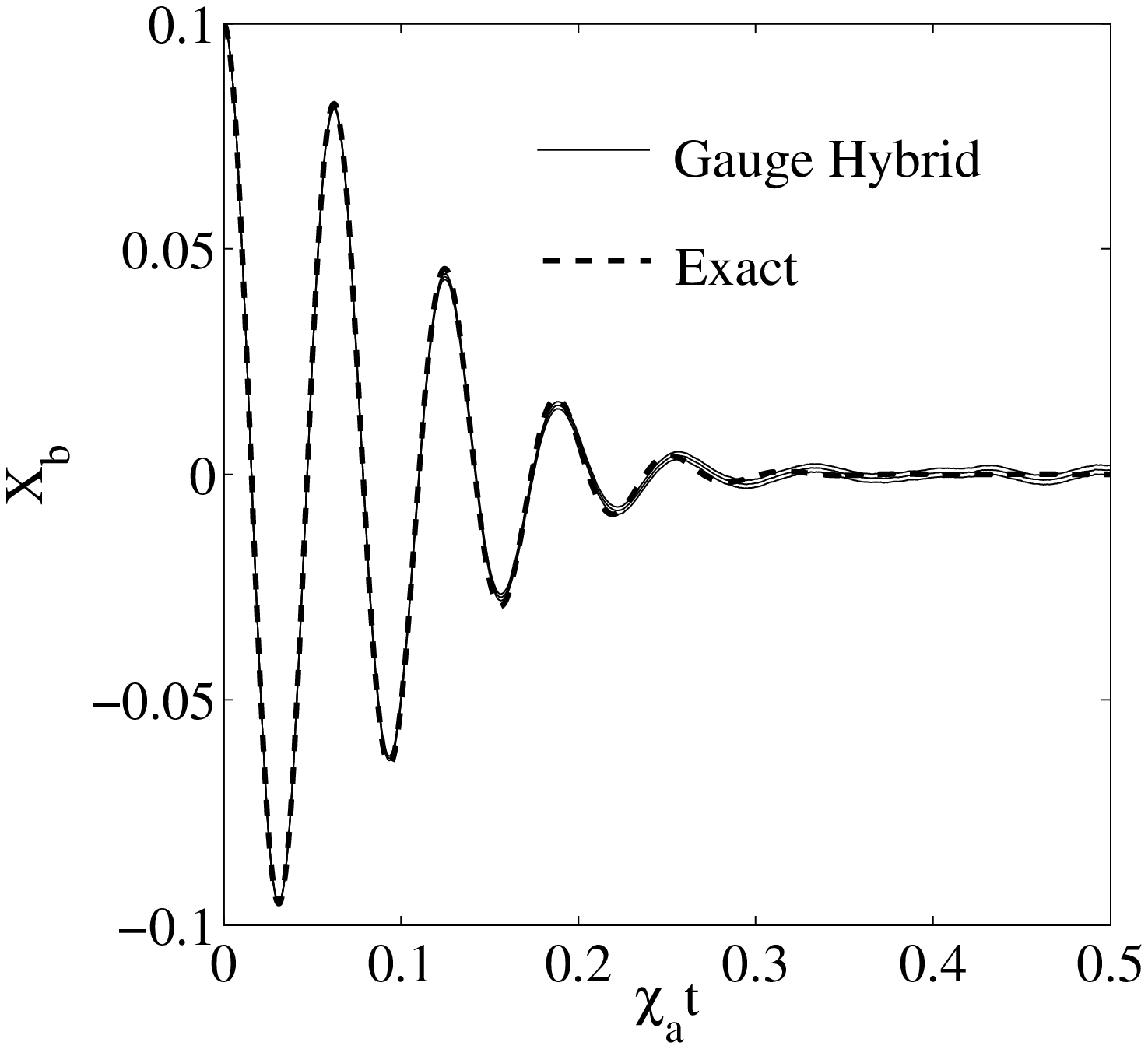}
\par\end{centering}

\noindent \begin{raggedright}
FIG. 3. X quadrature for mode $b$ \textit{vs} $\chi_{a}t$: coupled
anharmonic oscillators treated with the gauge Hybrid method. Plotted
are the ensemble average, the ensemble average~$\pm$~sampling error
estimate and the exact solution. Parameters: $\omega_{a}=\omega_{b}=0$,
$\chi_{a}=\chi_{b}=g=1,$ $N_{a0}=100$, $N_{b0}=0.01$. Number of
trajectories: 10,000.
\par\end{raggedright}

\end{figure}

\par\end{center}

When we simulated this same problem using the truncated Wigner method,
the results were nearly indistinguishable from Figures 2 and 3 (using
150,000 trajectories), so we do not display them here. With regard
to this first test, we have not yet established superiority of the
Hybrid method over the truncated Wigner, except to note that the Hybrid
method requires far fewer trajectories to attain a given accuracy.
In Section VIII we will explore a different region of parameter space
and in Section X we will calculate a higher-order moment in the same
system. In both cases, we will see results that show a clear distinction
between the methods.

We also simulated this problem with the positive-P method. Sampling
error rose to very large values at about $t=0.04,$ after just one
oscillation of the quadratures.

Analysis of the third-order derivative terms from the pure Wigner
calculation, similar to the above analysis for the Hybrid method,
shows terms that scale as $1/N_{b0}$ and so cannot be justifiably
neglected.

The mechanism at work in stabilizing the Hybrid simulation over limited
times is as follows. With this choice of gauge, the stochastic differential
equations keep the quantity $\alpha^{+}\alpha$ fixed, for each trajectory,
at its initial value. These values, selected by the stochastic Wigner
initial condition of the form (\ref{eq:W12},~\ref{eq:W13}), will
always be real and close to $N_{a0}.$ The quantity $\beta^{+}\beta$
starts at $N_{b0}$ then acquires an imaginary part, but its magnitude
is kept of order $N_{b0}$ over the simulation time.

Further inspection shows that the magnitudes of $\alpha$ and $\alpha^{+}$
will remain near $\sqrt{N_{a0}}$ while those of $\beta$ and $\beta^{+}$
remain of the order of $\sqrt{N_{b0}}$ over the simulation time.
(These estimates were used to justify our neglect of the third-order
derivative terms in equations (\ref{eq:7.1b},~\ref{eq:7.1b.1},~\ref{eq:7.1c}).)
So the drift terms are dominated by the factors of $\alpha^{+}\alpha$
and spiraling is negligible.

Over a short time $\Delta t$, the relative sizes of the drift and
diffusion increments, for $z$ one of the phase space variables, are
given by\[
\mbox{Drift:}\quad\Delta z\sim N_{a}\, z\,\Delta t\]
\[
\mbox{Diffusion:}\quad\Delta z\sim z\,\sqrt{\Delta t}\]
(with $\chi_{a}=\chi_{b}=g=1$). So diffusion is, in this example,
negligible compared to drift over the time scale of interest.

We calculated the quadrature $X_{a}$ in our model to longer times,
with results shown in Figure 4. (To obtain the qualitative features
rapidly, we used, in each case, a lower number of trajectories than
we used in our previous simulations.) The exact result showed a recurrence
centered on $t=\pi$. The gauge Hybrid method showed large sampling
error before that time, starting at about $t=2.5$. The truncated
Wigner method was also unable to predict this recurrence, showing
instead a quadrature remaining close to zero. Recalling that the pure
positive-P treatment suffered large sampling error after about $t=0.04,$
we see that the gauge Hybrid method extended the useful simulation
time by a factor of $60$.

\noindent \begin{center}
\begin{figure}[h]
\noindent \begin{centering}
\includegraphics[scale=0.4]{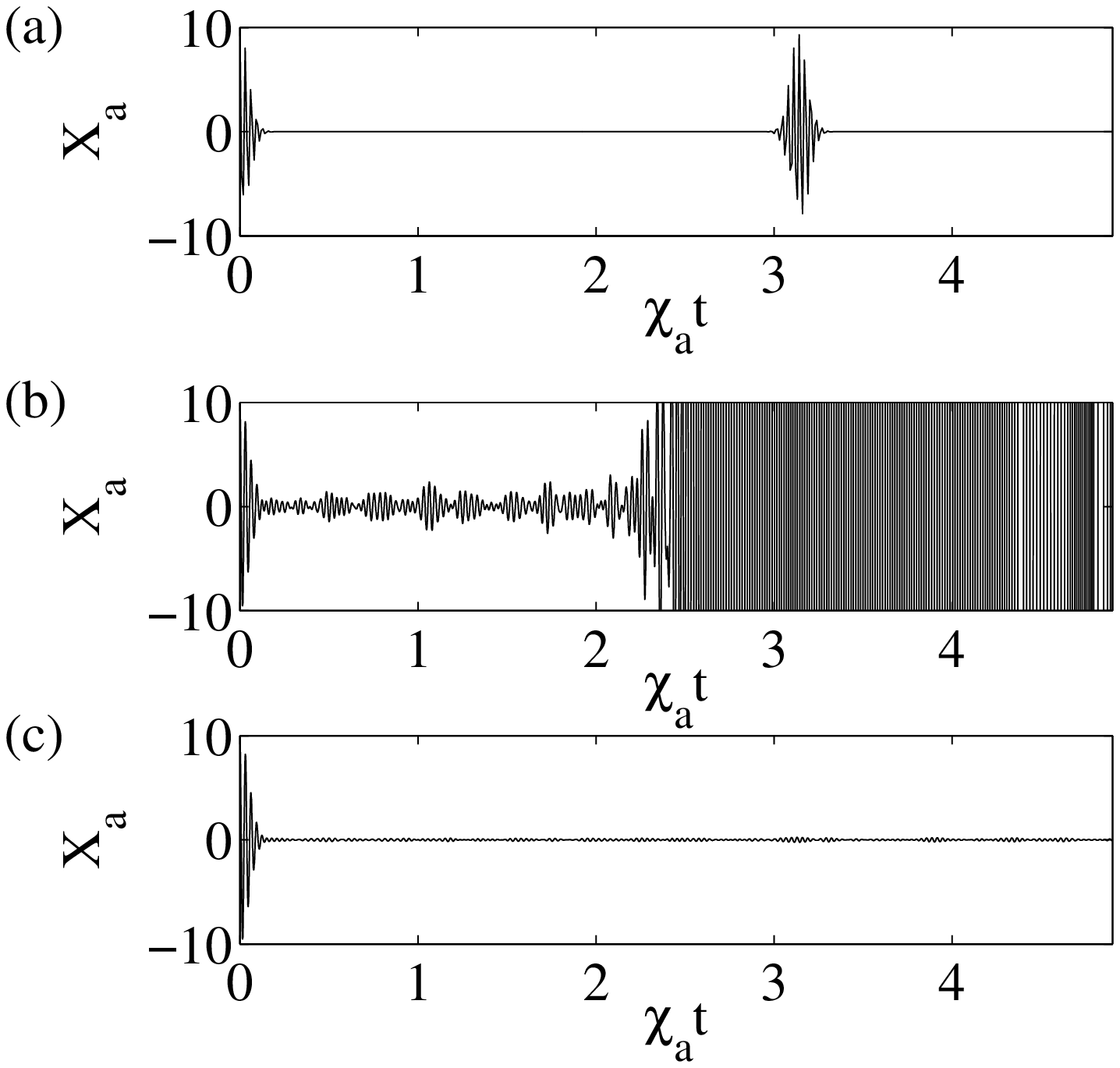}
\par\end{centering}

\noindent \begin{raggedright}
FIG. 4. X quadrature for mode $a$ calculated to longer (dimensionless)
times $\chi_{a}t$. (a) Exact result. (b) Gauge Hybrid result: 100
trajectories. (c) Truncated Wigner result: 10,000 trajectories. Parameters:
$\omega_{a}=\omega_{b}=0$, $\chi_{a}=\chi_{b}=g,$ $N_{a0}=100$,
$N_{b0}=0.01$.
\par\end{raggedright}

\end{figure}

\par\end{center}

\section{Weak Coupling}

In the previous example, both the gauge Hybrid method and the truncated
Wigner method are aided by the fact that the quadratures are strongly
damped before the neglect of terms (for both methods) and sampling
error growth (for the Hybrid method) can become important. We lowered
the mutual interaction strength between the two modes, relative to
$\chi_{a}=\chi_{b}$, by setting $g/\chi_{a}=0.0001.$ This greatly
extended the damping time for the $b$ mode, allowing us to see differences
in the predictions of the gauge Hybrid and truncated Wigner methods.

The results are shown in Figure 5.

\noindent \begin{center}
\begin{figure}[h]
\noindent \begin{centering}
\includegraphics[scale=0.4]{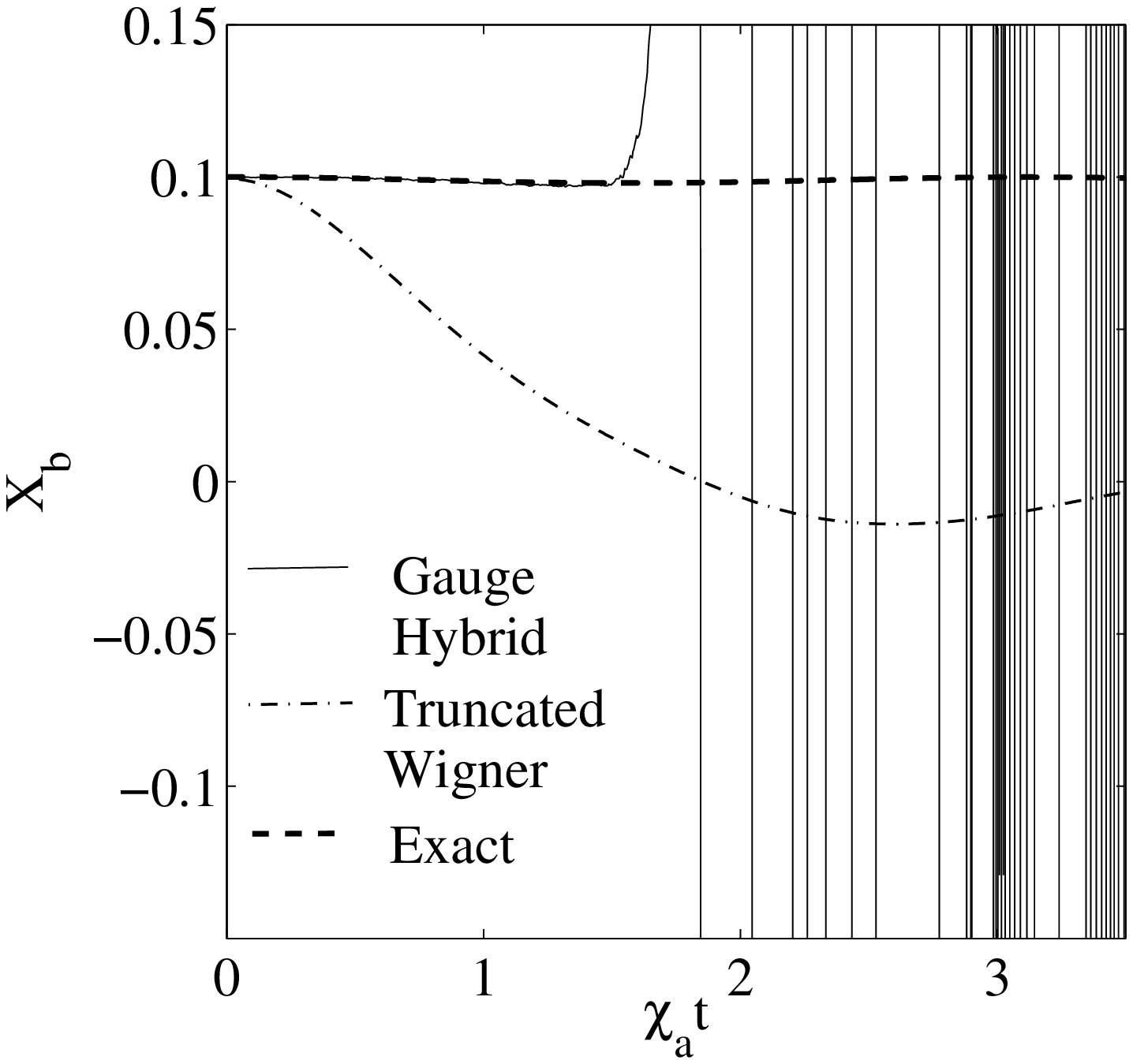}
\par\end{centering}

\noindent \begin{raggedright}
FIG. 5. Comparison of the gauge Hybrid and truncated Wigner methods
in predicting the X quadrature for the $b$ mode of the coupled anharmonic
oscillators, with weak coupling. Quadratures plotted against $\chi_{a}t.$
Gauge Hybrid: 10,000 trajectories. Truncated Wigner: 15,000 trajectories.
Parameters: $\omega_{a}=\omega_{b}=0$, $\chi_{a}=\chi_{b},$ $g/\chi_{a}=0.0001,$
$N_{a0}=100$, $N_{b0}=0.01$.
\par\end{raggedright}

\end{figure}

\par\end{center}

We see that the truncated Wigner method fails from $\chi_{a}t=0$,
consistent with our expectations for a system with a very lightly
occupied mode. The gauge Hybrid method performs well until about $\chi_{a}t=1.5$,
when it is overwhelmed by sampling error.

\section{Further Truncation}

In the examples we have seen so far, truncation of terms in the Hybrid
method has not prevented it from attaining excellent agreement with
the exact solutions at early times, even when dealing with a very
lightly occupied mode. The method is, however, clearly limited by
the growth of sampling error. In this section, we try a simple adjustment
to the equations to try to extend useful results to longer times.

When phase space distributions grow wide in unconstrained directions,
the trajectories sampling those distributions are widely spread and
the calculation of expectation values becomes a great numerical difficulty.
To understand the meaning of the widths in {}``unconstrained directions,''
we note that we could estimate the spread of our Hybrid distribution
by calculating all the stochastic averages $\langle\langle\alpha^{i}\alpha^{j}\rangle\rangle,$
where $\alpha^{i}$ is a real or imaginary part of $\vec{\alpha}$
(defined after equation (\ref{eq:6.6})). Some of the linear combinations
of these averages, such as $\langle\langle\alpha^{+}\alpha\rangle\rangle,$
are constrained to approach physical predictions as the number of
trajectories grows large. Widening of the distribution in the other
directions will increase sampling error, but may be reduced using
the gauge freedoms of doubled phase-space representations, or other
methods.

Spiraling of the drift trajectories is one source of spreading that
we have identified, and that we have already partially controlled
using our choice of gauge. For our gauge SDEs, the quantity $\alpha^{+}\alpha$
remains completely real for all times, and thus does not cause spiraling
in the drift equations. Not so the quantity $\beta^{+}\beta,$ which
starts with a purely real value but can immediately develop an imaginary
part from the influence of the noise terms.

We tried a further truncation of our gauge Hybrid equations, making
the replacement\begin{equation}
\beta^{+}\beta\rightarrow\mathrm{Re}(\beta^{+}\beta).\label{eq:T.1}\end{equation}
In future applications, if $\alpha^{+}\alpha$ is not constrained,
we propose to also try the truncation\begin{equation}
\alpha^{+}\alpha\rightarrow\mathrm{Re}(\alpha^{+}\alpha).\label{eq:T.2}\end{equation}

\begin{center}

\par\end{center}

We saw good short-time behavior from this truncated Hybrid method,
equaling that of all the other methods. At longer times the method
was unable to predict the recurrence, showing quadratures staying
close to zero. But the sampling error remained at a manageable level
to $t=5.0.$ In future work, we will investigated whether this somewhat
\textit{ad hoc} truncation can be used as a simple way to extend simulations
to longer times without incurring excessive systematic error.

\section{Test Case: Quantum Nondemolition Measurement}

Our first test case showed the Hybrid method---with a diffusion gauge
choice and with a further truncation---able to successfully simulate
an interacting system beyond the time at which the positive-P method
became unusable. But the Wigner method was able to give equally good
results on the same system. (A distinction was found in the weak coupling
case.) Here we investigate a different observable---a higher order
moment---in the same system, and find the results more sensitive to
the choice of method.

The concept of quantum nondemolition measurements \citep{Caves1980,Holland1990}
arose from the need for a way to measure the very small displacements
of a gravitational wave detector that are expected to occur from the
passage of a gravitational wave. Repeated measurements of position,
to high accuracy, would be required to distinguish the signal from
other effects. Quantum mechanics sets limits on schemes to measure
those small displacements. Measurement of a position observable with
a finite uncertainty may produce a state in which the uncertainty
in position grows after the measurement. At later times, when another
measurement of position is performed, the uncertainty would be larger
that the desired maximum.

Instead, measurement of a conserved observable, such as the momentum
of a free particle, can be repeated an arbitrary number of times without
causing the uncertainty to increase. The quantum nondemolition (QND)
measurement scheme involves choosing an appropriate conserved observable
(in a probe beam) that can give information about the signal of interest
after the signal and probe interact.

A QND scheme can be constructed from our model of interacting anharmonic
oscillators \citep{Milburn1983a}. We suppose that the bosons in question
are now photons, and that they can interact with each other in a suitable
nonlinear medium, such that our number-conserving interaction Hamiltonian
gives a toy model of the dynamics. Of course a fuller description
of the dynamics would involve propagation in space, dispersion and
other factors \citep{Dong2008}. A lightly occupied signal beam and
a highly occupied probe beam interact in the medium. Phase information
will be exchanged between them, while their individual number distributions
are conserved.

In one QND scheme, the conserved QND observable is taken as the photon
number, $\hat{N}_{a},$ in the highly occupied probe beam. The signal
is the phase quadrature of the lightly occupied beam, $\hat{Y}_{b}=-\frac{i}{2}(\hat{b}^{\dagger}-\hat{b}).$

We suppose that the interaction between signal and probe lasts only
for a short time, as would be the case for two short pulses interacting
in an optical fiber. We make the interaction cease when the magnitude
of the correlation function reaches its first maximum. This means
that we use the previous Hamiltonian of equation (\ref{eq:7.1}),
except with $g=1$ for $t<t_{0}$, and $g=0$ for $t>t_{0}$, where
$t_{0}=0.1$ in our example.

We calculate the correlation function between probe and signal, a
measure of the potential success of the measurement scheme:

\begin{equation}
C(N_{a},Y_{b})=\frac{\langle\hat{N}_{a}\hat{Y}_{b}\rangle-\langle\hat{N}_{a}\rangle\langle\hat{Y}_{b}\rangle}{V^{\frac{1}{2}}(N_{a})V^{\frac{1}{2}}(Y_{b})},\label{eq:9.1}\end{equation}
where\begin{equation}
V(\hat{\Omega})=\langle\hat{\Omega}^{2}\rangle-\langle\hat{\Omega}\rangle^{2}\label{eq:9.2}\end{equation}
is the variance of an operator $\hat{\Omega}.$ We set $\omega_{a}=0$
for convenience, since this will remove a high frequency variation
from our expectation value. Likewise, we set $\omega_{b}=-N_{a0}g$
to obtain a slowly varying expectation value. This latter choice is
equivalent to a particular choice of local oscillator frequency in
the homodyne detection of $Y_{b}.$

Figure 6 shows the correlation function calculated with two different
phase-space methods and compared to the exact result. The gauge Hybrid
method shows excellent agreement with the exact result. In contrast,
the truncation of the Wigner method evidently removes terms that are
needed to correctly predict the correlation function at times after
the interaction ceases. The tendency of the truncated Wigner method
to give worse results when predicting higher-order moments was investigated
by Drummond \textit{et al} \citep{Drummond1995}.

\noindent \begin{center}
\begin{figure}[h]
\begin{centering}
\includegraphics[scale=0.4]{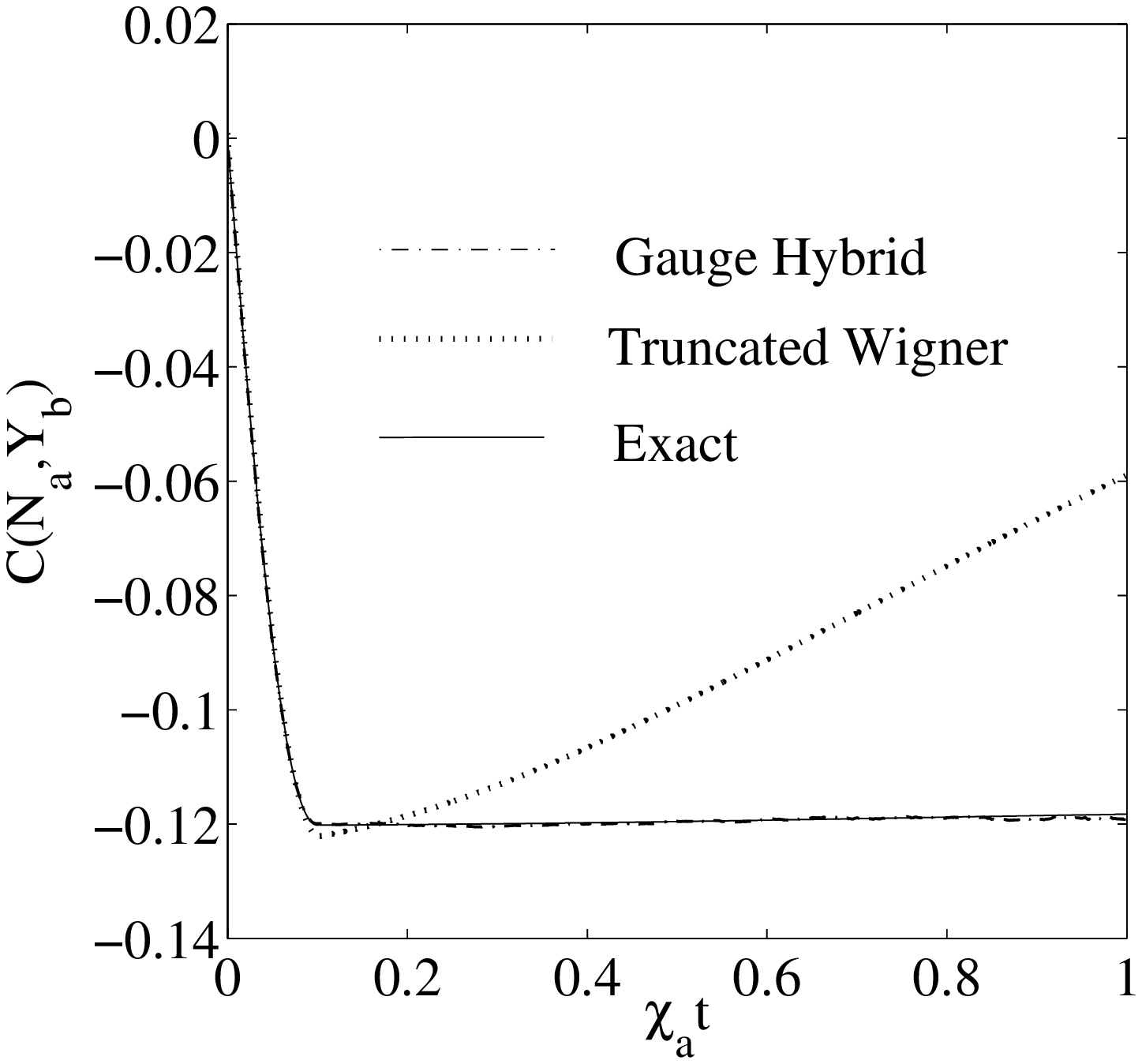}
\par\end{centering}

\noindent \begin{raggedright}
FIG. 6. Comparison of methods for determining the correlation between
$N_{a}$ and $Y_{b}$ for a QND scheme, \textit{vs} $\chi_{a}t$.
Results are shown for the Hybrid method with a diffusion gauge (50,000
trajectories) and the truncated Wigner method (50,000 trajectories),
compared to the exact result. Parameters: $N_{a0}=100$, $N_{b0}=0.01$,
$\omega_{a}=0$, $\omega_{b}=-N_{a0}g$, $\chi_{a}=\chi_{b}$, $g/\chi_{a}=1$
for $\chi_{a}t<0.1,$ $g=0$ for $\chi_{a}t>0.1.$
\par\end{raggedright}

\end{figure}

\par\end{center}

\section{Conclusions}

We have shown that, for a stochastic phase-space treatment of a multimode
system, it is possible to use the doubled Wigner representation for
some modes and the positive-P representation for the remainder. We
tested our method on a system of two coupled anharmonic oscillators,
one with a mean occupation that remained at 100, the other with a
mean occupation of 0.01. The method was able to simulate the evolution
of quadrature expectation values for times far beyond where the positive-P
method suffers a rapid growth of sampling error. Results were in excellent
agreement with the exact solution.

While the truncated Wigner method performed as well as the Hybrid
method when calculating these quadrature observables (over a finite
time), for the calculation of a higher order moment corresponding
to a QND experiment there was a very clear advantage of the Hybrid
over the truncated Wigner. The latter results contained a large systematic
error, while the Hybrid result was in excellent agreement with the
exact result.

At least as applied to this system with a small number of modes, the
Hybrid method was able to delay the onset of rapid sampling error
growth by a factor of $60$ compared to the positive-P method. Further
investigations will focus on many-mode systems to see whether these
advantages over the earlier methods can be maintained. 

It is interesting to note here that our results show that a very natural
application of the Hybrid method is to systems of two different types
of particle with interactions that conserve individual species numbers.
This presents a natural framework to investigate quantum Brownian
motion, which will be treated in subsequent work.

\begin{acknowledgments}
This research was supported by the Australian Research Council Centre
of Excellence for Quantum-Atom Optics. S.H. holds an Australian Postgraduate
Award.
\end{acknowledgments}
  \renewcommand{\theequation}{A-\arabic{equation}}   

\section*{APPENDIX: EXACT SOLUTIONS}

The Hamiltonian\begin{equation}
\hat{H}=\omega_{a}\hat{a}^{\dagger}\hat{a}+\chi_{a}\hat{a}^{\dagger}\hat{a}^{\dagger}\hat{a}\hat{a}+\omega_{b}\hat{b}^{\dagger}\hat{b}+\chi_{b}\hat{b}^{\dagger}\hat{b}^{\dagger}\hat{b}\hat{b}+g\hat{a}^{\dagger}\hat{a}\hat{b}^{\dagger}\hat{b},\label{eq:A1}\end{equation}
describing two coupled anharmonic oscillators, can be written just
in terms of the number operators, $\hat{N}_{a}=\hat{a}^{\dagger}\hat{a}$
and $\hat{N}_{b}=\hat{b}^{\dagger}\hat{b}$, as\begin{eqnarray}
\hat{H} & = & \omega_{a}\hat{N}_{a}+\omega_{b}\hat{N}_{b}+\chi_{a}(\hat{N}_{a}^{2}-\hat{N}_{a})\label{eq:A2}\\
 &  & \quad+\chi_{b}(\hat{N}_{b}^{2}-\hat{N}_{b})+g\hat{N}_{a}\hat{N}_{b}.\nonumber \end{eqnarray}
So the number states\begin{equation}
|n_{a}n_{b}\rangle=\frac{\hat{a}^{\dagger n_{a}}}{\sqrt{n_{a}!}}\frac{\hat{a}^{\dagger n_{b}}}{\sqrt{n_{b}!}}|0\rangle\label{eq:A3}\end{equation}
are eigenvectors of the Hamiltonian with eigenvalues\begin{eqnarray}
E(n_{a},n_{b}) & = & \omega_{a}n_{a}+\omega_{b}n_{b}+\chi_{a}(n_{a}^{2}-n_{a})\nonumber \\
 &  & \quad+\chi_{b}(n_{b}^{2}-n_{b})+gn_{a}n_{b}.\label{eq:A4}\end{eqnarray}

We consider an initial state that is a coherent superposition of the
number states (\ref{eq:A3}) of the form\begin{equation}
|\gamma_{a}\gamma_{b}\rangle=\sum_{n_{a}=1}^{\infty}e^{-\frac{1}{2}|\gamma_{a}|^{2}}\frac{\gamma_{a}^{n_{a}}}{\sqrt{n_{a}!}}\sum_{n_{b}=1}^{\infty}e^{-\frac{1}{2}|\gamma_{b}|^{2}}\frac{\gamma_{b}^{n_{b}}}{\sqrt{n_{b}!}}|n_{a}n_{b}\rangle,\label{eq:A5}\end{equation}
where $\gamma_{a}=\sqrt{N_{a0}}$, $\gamma_{b}=\sqrt{N_{b0}}$ and
$N_{a0}$ and $N_{b0}$ are the average occupations of the modes.

We are interested in observables, $\Omega,$ that are simple combinations
of a small number of creation and/or annihilation operators. These
have simple matrix elements, $\langle n'_{a}n'_{b}|\hat{\Omega}|n_{a}n_{b}\rangle,$
between the number eigenvectors. All terms will be proportional to
Kronecker deltas of the form $\delta_{n'_{a},n_{a}+m_{a}}\delta_{n'_{b},n_{b}+m_{b}},$
for $m_{a}$ and $m_{b}$ integers. Then the expectation value of
such an operator in the state vector produced by time evolution of
(\ref{eq:A5}) will always reduce to a double sum (over $n_{a}$ and
$n_{b}$), with the unitary time evolution producing a known phase
factor inside the sum.

The time-dependent expectation values of the quadrature operators
$X_{a}=\frac{1}{2}(a+a^{\dagger}),$ $Y_{a}=\frac{1}{2i}(a-a^{\dagger}),$
$X_{b}=\frac{1}{2}(b+b^{\dagger})$ and $Y_{b}=\frac{1}{2i}(b-b^{\dagger})$,
for the initial state (\ref{eq:A5}), can then be evaluated as sums
over $n_{a}$ and $n_{b}.$ We find the results\begin{eqnarray}
\langle X_{a}(t)\rangle & = & \sqrt{N_{a0}}e^{-\{N_{a0}(1-\cos2\chi_{a}t)+N_{b0}(1-\cos gt)\}}\quad\quad\label{eq:A6}\\
 &  & \times\cos\{\omega_{a}t+N_{a0}\sin2\chi_{a}t+N_{b0}\sin gt\},\nonumber \end{eqnarray}
\begin{eqnarray}
\langle Y_{a}(t)\rangle & = & -\sqrt{N_{a0}}e^{-\{N_{a0}(1-\cos2\chi_{a}t)+N_{b0}(1-\cos gt)\}}\quad\quad\nonumber \\
 &  & \times\sin\{\omega_{a}t+N_{a0}\sin2\chi_{a}t+N_{b0}\sin gt\}.\label{eq:A6a}\end{eqnarray}
For $\langle X_{b}(t)\rangle$ and $\langle Y_{b}(t)\rangle,$ we
make the replacement $a\leftrightarrow b$ in expressions (\ref{eq:A6})
and (\ref{eq:A6a}), respectively.

Our model of a QND measurement has the feature that the coupling strength,
$g,$ is constant up to a time, $\tau,$ and vanishes after that.
This is meant to model two light pulses that cease to interact after
they no longer overlap within an optical fiber. The evolution operator
for the resulting time-dependent Hamiltonian is $\hat{U}(t)=\exp-\int_{0}^{t}\hat{H}(t')dt'.$
For times up to $\tau,$ we evaluate the following: \begin{eqnarray}
\langle\hat{N}_{a}\hat{Y}_{b}(t)\rangle & = & -N_{a}\sqrt{N_{b}}e^{-N_{a0}(1-\cos gt)}\nonumber \\
 &  & \times e^{-N_{b0}(1-\cos2\chi_{b}t)}\label{eq:A7}\\
 &  & \times\sin\{(\omega_{b}+g)t+N_{a0}\sin gt+N_{b0}\sin2\chi_{b}t\},\nonumber \end{eqnarray}
\begin{eqnarray}
V(\hat{Y}_{b}) & = & \langle\hat{Y}_{b}^{2}\rangle-\langle\hat{Y}_{b}\rangle^{2}\nonumber \\
 & = & -\frac{1}{2}N_{b0}e^{-\{N_{a}(1-\cos2gt)+N_{b}(1-\cos4\chi_{b}t)\}}\times\nonumber \\
 &  & \times\cos\{2(\omega_{b}+\chi_{b})t+N_{a}\sin2gt+N_{b}\sin4\chi_{b}t\}\nonumber \\
 &  & -N_{b}e^{-2\{N_{a}(1-\cos gt)+N_{b}(1-\cos2\chi_{b}t)\}}\times\label{eq:A8}\\
 &  & \times\sin^{2}\{\omega_{b}t+N_{a}\sin gt+N_{b}\sin2\chi_{b}t\}\nonumber \\
 &  & +\frac{1}{4}+\frac{1}{2}N_{b0}.\nonumber \end{eqnarray}
For times beyond $\tau$, we make the replacement $gt\rightarrow g\tau$
in (\ref{eq:A7}) and (\ref{eq:A8}). This rule applies also to the
expectation values (\ref{eq:A6}) and (\ref{eq:A6a}) for $t>\tau$
in this QND scheme.

Finally we note that the individual particle numbers are conserved
under this Hamiltonian, and the coherent state initial conditions
(\ref{eq:A5}) give\begin{equation}
\langle\hat{N}_{a}\rangle=N_{a0},\quad\langle\hat{N}_{b}\rangle=N_{b0},\label{eq:A9}\end{equation}
\begin{equation}
V(\hat{N}_{a})=N_{a0},\quad V(\hat{N}_{b})=N_{b0}.\label{eq:A10}\end{equation}

\bibliographystyle{prsty}
\bibliography{/data/hoffmann/References/References}

\end{document}